\documentclass[twocolumn,aps,floatfix,showpacs]{revtex4}
\usepackage{pstricks}
\usepackage{pst-coil}
\usepackage{amssymb}
\usepackage{graphics,epsf}
\newcommand{\be}{\begin{equation}}
\newcommand{\ee}{\end{equation}}
\begin{document}
\title{Naked singularities in cylindrical collapse of counter-rotating dust shells.}
\author{Brien C Nolan\footnote{brien.nolan@dcu.ie}}
\affiliation{School of Mathematical Sciences, Dublin City
University, Glasnevin, Dublin 9, Ireland.}
\date{\today}
\begin{abstract}
Solutions describing the gravitational collapse of asymptotically
flat cylindrical and prolate shells of (null) dust are shown to
admit globally naked singularities.
\end{abstract}
\maketitle
We consider two models of cylindrical
collapse, both based on the cylindrically symmetric line element
\be ds^2=-\exp(-\alpha(v))dudv+dz^2+r^2(u,v)d\phi^2.
\label{linel}\ee $u,v$ are null co-ordinates and we take
$r=(v-u)/2\geq 0$. Then $u,v$ label outgoing and ingoing null
hypersurfaces respectively; we take both $u$ and $v$ to increase
into the future. The energy-momentum tensor is in the form of null
dust with Ricci tensor given by
\[ R_{ab}=-\frac{\alpha^\prime(v)}{4r}k_ak_b,\qquad k_a=\nabla_av.\]
Then the strong, weak and dominant energy conditions are all
equivalent to \be\alpha^\prime\leq 0,\label{econ}\ee which we
assume henceforth. The matter content of the space-time may be
described as infalling null dust. There is a curvature singularity
along $r=0$. Since the solution is Petrov type $N$ with a pure
radiation energy-momentum tensor, all curvature invariants vanish.
It has been shown in \cite{letwang} that this is a parallel
propagated curvature singularity. Additionally, scalars such as
$R_{ab}w^aw^b$ diverge at $r=0$ for any unit time-like vector
$w^a$. At fixed $z$ on the singularity, we have
\[ ds^2=-e^{-\alpha(v)}dv^2,\]
so the singularity is time-like.

The causal nature of the singularity is further elucidated by
studying the null geodesics in the 2-space $z=$constant,
$\phi=$constant. The geodesic equations yield (for future-pointing
geodesics)
\begin{eqnarray}
{\dot v}&=&k^2\exp(\alpha(v)),\label{g1}\\
u&=&u_0+u_1\tau,\label{g2}
\end{eqnarray}
where the over-dot represents differentiation with respect to the
affine parameter $\tau$, $u_1\geq 0$ and $k$ are constants. The
null condition is $u_1k^2=0$; both constants cannot vanish
simultaneously.

If $k=0$, then $u_1>0$. We see that $u\to-\infty,r\to+\infty$ as
$\tau\to-\infty$. As $\tau$ increases, $r$ decreases to $0$ at
parameter time $\tau=(v_0-u_0)/u_1>0$. So these geodesics are past
complete and terminate in the future at the singularity $r=0$ in
finite parameter time.

If $u_1=0$, then $k^2>0$. We calculate
\[{\ddot{v}}=\alpha^\prime(v){\dot{v}}^2,\]
so $\tau\to v(\tau)$ is increasing and concave (according to the
energy condition (\ref{econ})). Hence $v$ cannot reach $+\infty$
in finite parameter time. Either $v\to\infty$ as $\tau\to\infty$,
or $v$ is bounded above. In the latter case, monotonicity implies
that $v$ has a limit $v_F$ as $\tau$ increases. Then either $v\to
v_F$ as $\tau\to\infty$, or $v=v_F$ at some $\tau=\tau_F$ with
${\dot{v}}(\tau_F)=0$. This implies $\alpha(v_F)=-\infty$, and by
monotonicity, $\alpha^\prime(v_F)=-\infty$. In this case, these is
a singularity along $v=v_F$. In the other cases, the geodesics are
future complete. Future null infinity ${\cal{J}}^+$ is at $v=v_F$
or  $v=+\infty$  depending on whether or not $v$ is bounded above
as $\tau\to\infty$. In the former case, $r$ has a finite limit at
${\cal{J}}^+$, which will be different along different geodesics.

$v$ decreases as $\tau$ decreases. Either $r$ decreases to $0$, or
the geodesic meets a singularity at $v=v_P$ whereat
${\dot{v}}\to\infty$. This occurs if and only if
$\alpha(v_P)=+\infty$. Again, monotonicity implies that
$\alpha^\prime(v_P)=-\infty$.

So there are four different possible global structures depending
on whether or not there exist values $v_P, v_F$ for which
$\alpha(v_{P,F})=+\infty,-\infty$. The conformal diagrams are
given in Figures 1-4. Note that when ${\cal{J}}^{\pm}$ exists and
corresponds to $r\to\infty$, the space-time is asymptotically flat
at that surface in the following sense: all superenergy terms
\cite{jose} such as $R_{ab}w^aw^b, C_{abcd}w^aw^bw^cw^d$ ($w^a$
null or unit time-like) vanish in the limit as the surface is
approached {\em for fixed values of $z$}. Along any time-like
geodesic, parallel propagated components of the Riemann tensor
will also vanish in this limit \cite{letwang}. Since the
space-times represent infinite cylinders, they cannot be
asymptotically flat in directions along which $z$ becomes
infinite. This feature can be removed; see below.

\begin{figure}[!htb]
\centerline{\def\epsfsize#1#2{1 #1}\epsffile{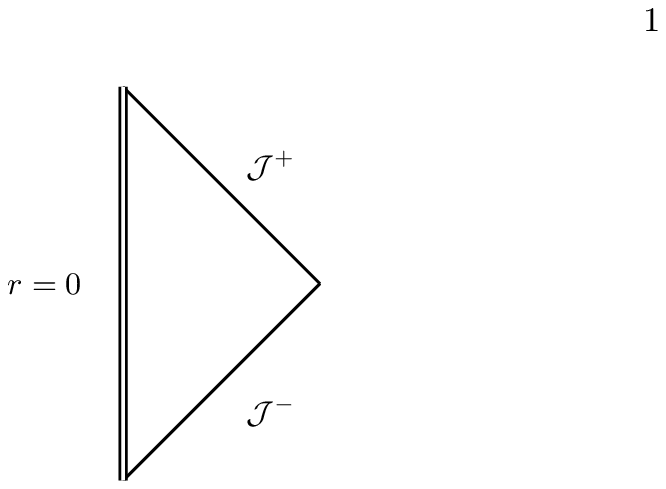}}
\caption{\label{figure1} Conformal diagram for the case when
$\alpha$ is bounded on $\mathbb{R}$. $v$ lies in the range
$(-\infty,+\infty)$. Singularities are indicated with a double
line.}
\end{figure}

\begin{figure}[!htb]
\centerline{\def\epsfsize#1#2{1 #1}\epsffile{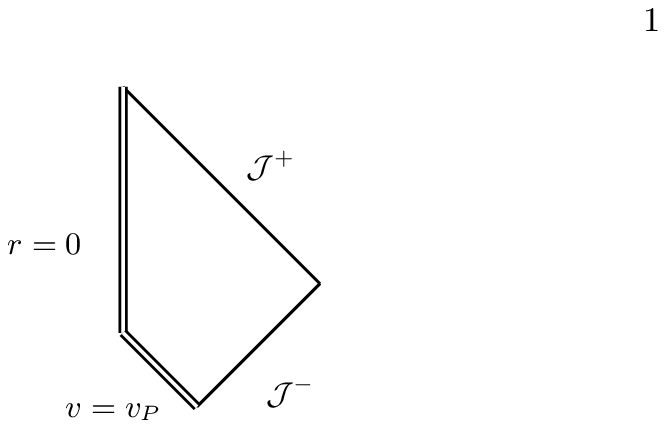}}
\caption{\label{figure2} Conformal diagram for the case when
$\alpha$ is bounded below on $\mathbb{R}$, but there exists
$v_P\in \mathbb{R}$ such that  $\alpha\to+\infty$ as $v\downarrow
v_P$. $v$ lies in the range $(v_P,+\infty)$.}
\end{figure}

\begin{figure}[!htb]
\centerline{\def\epsfsize#1#2{1 #1}\epsffile{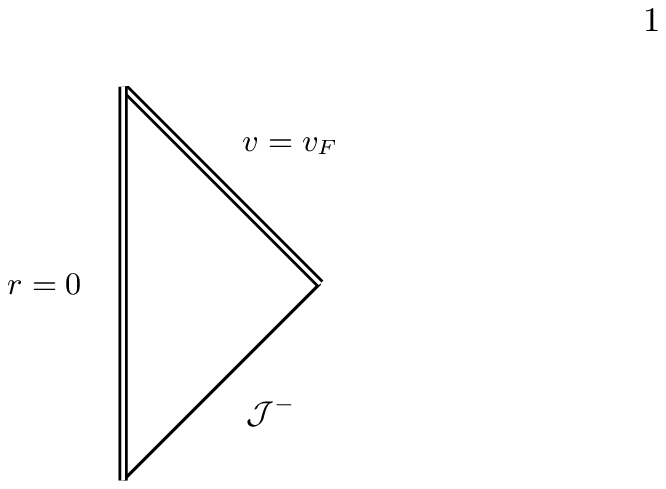}}
\caption{\label{figure3} Conformal diagram for the case when
$\alpha$ is bounded above on $\mathbb{R}$, but there exists
$v_F\in \mathbb{R}$ such that  $\alpha\to-\infty$ as $v\uparrow
v_F$. $v$ lies in the range $(-\infty,v_F)$.}
\end{figure}

\begin{figure}[!htb]
\centerline{\def\epsfsize#1#2{1 #1}\epsffile{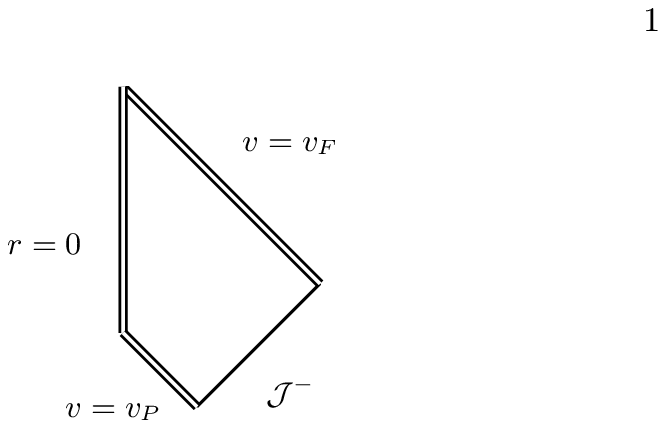}}
\caption{\label{figure4}Conformal diagram for the case when there
exists $v_P,v_F\in \mathbb{R}$ such that  $\alpha\to+\infty$ as
$v\downarrow v_P$ and $\alpha\to-\infty$ as $v\uparrow v_F$. $v$
lies in the range $(v_P,v_F)$.}
\end{figure}

We note that to obtain space-times filled with {\em outgoing} null
dust, we define $U=-v,V=-u,\beta(U)=\alpha(v)$ and re-designate
the future direction as the reverse of that used above. The line
element is \be
ds^2=-\exp(-\beta(U))dUdV+dz^2+R^2d\phi^2,\label{linel2}\ee where
$R=(V-U)/2$. The Ricci tensor is given by \[
R_{ab}=\frac{\beta^\prime(U)}{4r}\nabla_aU\nabla_bU,\] and the
energy condition now reads \be \beta^\prime(U)\geq
0.\label{econ2}\ee The relevant conformal diagrams are obtained by
inverting those of Figures 1-4.

In order to make the connection with cosmic censorship, i.e.\ to
study collapse for an initially regular configuration, we replace
the interior of a past light cone of a point on $r=0$ with a
portion of Minkowski space. The resulting space-times are
asymptotically flat and evolve from regular initial data. They may
be considered the cylindrical version of Vaidya space-time.
Writing the line element of Minkowski space-time in the
cylindrical form
\[ ds^2=-d{\bar u}d{\bar v}+d{\bar z}^2+\frac14({\bar v}-{\bar
u})^2d{\bar\phi}^2,\] we obtain a smooth matching across an
ingoing null hypersurface by taking ${\bar x}^a=x^a$, matching
across $v=0$ (a translation may be required to guarantee that $0$
lies in the domain of $\alpha$ and that $v_P<0<v_F$ if such exist)
and using the boundary condition $\alpha(0)=0$ \cite{BI}. We paste
the portion $v<0$ of Minkowski space-time to the portion $v\geq 0$
of the space-time with line element (\ref{linel}).

Two possibilities arise and are shown in Figures 5 and 6
respective. Both contain locally naked singularities. Figure 5
corresponds to space-times modelling the gravitational collapse of
cylindrical null dust which are asymptotically flat at fixed $z$
and which arise from regular initial data. The singularity in this
case is globally naked.

\begin{figure}[!htb]
\centerline{\def\epsfsize#1#2{1 #1}\epsffile{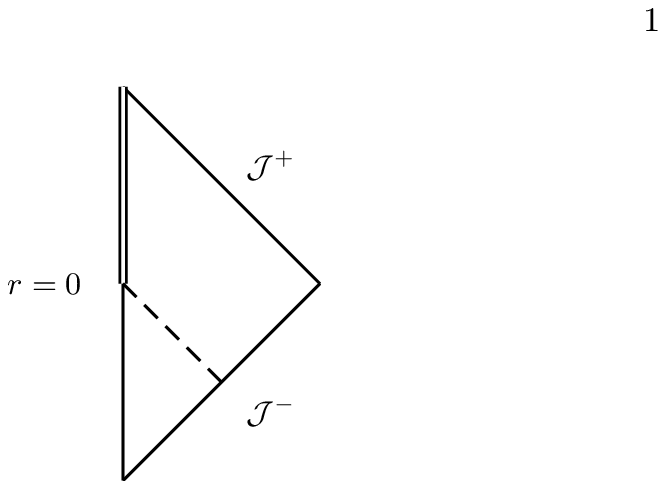}}
\caption{\label{figure5} Conformal diagram in the $u-v$ plane for
the case when $\alpha$ is bounded below on $\mathbb{R}$. The
dashed line is $v=0$; the portion to the past of $v=0$ is flat.}
\end{figure}

\begin{figure}[!htb]
\centerline{\def\epsfsize#1#2{1 #1}\epsffile{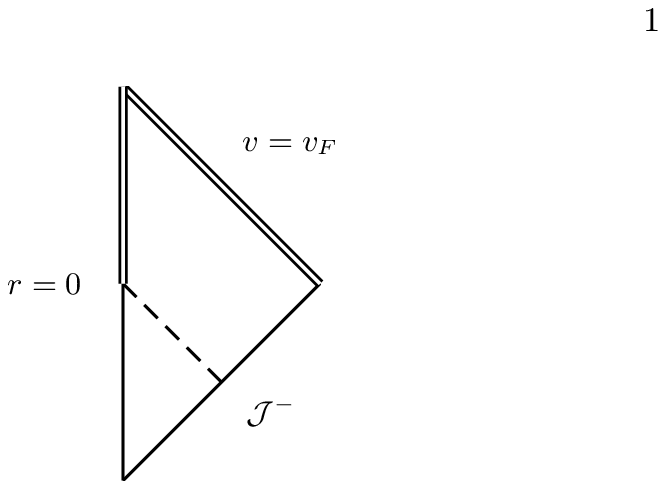}}
\caption{\label{figure6} Conformal diagram for the case when
$\alpha\to+\infty$ as $v\uparrow v_F>0$. The dashed line is $v=0$;
the portion to the past of $v=0$ is flat.}
\end{figure}

A slightly different model has been studied in \cite{PW},
\cite{GJ} and \cite{Gleiser}. Here, the exterior region is taken
to be filled with {\em outgoing} null dust, and is matched across
a time-like shell of (ordinary) dust to a flat interior. One can
interpret this as a collapsing cylindrical shell emitting both
gravitational radiation (the solution is Petrov type $N$) and
massless particles. It was shown in \cite{GJ} and \cite{Gleiser}
that this shell can be constructed in such a manner that the
collapse proceeds to $r=0$ without any trapped or marginally
trapped surfaces appearing either on the shell or in the exterior
geometry. This leads one to suspect that the resulting singularity
may be naked; the present results confirm this. The shell is
constructed as follows.  The metric inside the shell is taken to
be flat;
\[ ds^2 = -dt^2+dr^2+dz^2+r^2d\phi^2.\] The exterior geometry is given by
(\ref{linel2}). Define $T$ by $U=T-R$, $V=T+R$, and take the shell
to be at $R=R_0(T)=r_0(t)$. Demanding continuity of the metric
yields a relation for $t$ in terms of $T$. The surface
energy-momentum tensor of the shell general takes the form of an
imperfect fluid; following \cite{PW}, we impose the condition that
there are stresses only in the azimuthal direction. This choice of
matter admits the interpretation of a thin shell of dust particles
in which counter-rotation of one half of the particles results in
zero net angular momentum, and is referred to as a ``shell of
counter-rotating dust''\cite{AT,PW,Gleiser}. The field equations
for the shell yield the following differential equation for
$R_0(T)$ which includes the terms $\beta(U_0),\beta^\prime(U_0)$,
where $U_0=T-R_0(T)$: \be R_0^{\prime\prime}=(1-{R_0^\prime}^2)
\left\{\frac{\triangle^{1/2}}{R_0}+\frac{R_0^\prime-\triangle^{1/2}}{\triangle^{1/2}-1}
\frac{\beta^\prime}{2}(1-R_0^\prime)\right\},\label{mainr}\ee
where
\[\triangle={R_0^\prime}^2+e^{-beta}(1-{R_0^\prime}^2).\]
(An erroneous version of this equation was given in \cite{PW} and
was corrected in \cite{Gleiser}.)

The solutions studied in \cite{PW}, \cite{GJ} and \cite{Gleiser}
are similar to the following simple example. For our present
purposes, it suffices to show that there exists a function
$\beta(U)$ and a solution $R_0(T)$ of (\ref{mainr}) with the
following properties: $R_0(T)$ decreases monotonically from a
finite positive value to $R_0=0$ in finite proper time;
$|R_0^\prime|<1$; $\beta^\prime(U)\geq 0$; $\triangle>0$ along the
solution; a suitable energy condition is satisfied by the shell.
In order to do this, we make the choice $R_0(T)=a-bT$ where
$a,b>0$ and $b<1$. Then (\ref{mainr}) may be written as the
following equation for $y(\tau):=\beta(T-R_0(T))$ where
$\tau=-\ln|a-bT|$:
\be\frac{dy}{d\tau}=\frac{2}{b}\frac{\triangle-\triangle^{1/2}}{b+\triangle^{1/2}}.\label{mainy}\ee
The right hand side here is a $C^1$ function of $y$, and so this
equation admits a unique solution through any point
$(\tau_0,y_0)$. If we choose data with $y<0$, we see that $y$
increases towards $y=0$ asymptotically as $\tau\to\infty$. Such a
solution cannot reach $y=0$ in a finite time since $y\equiv 0$ is
the unique solution through any point $(\tau,y)=(\tau_0,0)$. In
other words, there exist solutions of (\ref{mainr}) for which
$R_0$ decreases to zero in finite time $T=a/b$, with $\beta$
satisfying $\beta_0\leq\beta\leq 0$ on $T\in[0,a/b]$. The shell
has density $\rho$ and azimuthal stress $p_\phi$ satisfying
\[\rho=p_\phi=e^{\beta/2}(1-{R_0^\prime}^2)^{-1/2}R_0^{-1}(\triangle^{1/2}-1),\]
which is clearly positive for the solution constructed, and the
other conditions mentioned above are clearly satisfied. We note
that Gleiser's general analysis shows that there are numerous
possibilities which lead to the outcome exemplified by this case
\cite{Gleiser}.

To see that the resulting singularity is globally naked, we note
that since  $\beta(U_0)\leq 0$ during the collapse, $\beta$ cannot
diverge to $+\infty$ before the shell undergoes complete collapse
and hence the conformal diagram for the matched space-time must be
of the form shown in either Figure 7 or Figure 8. When there is no
future singularity $U_F$, i.e.\ when Figure 7 applies, the
space-times give more examples of cylindrical collapse resulting
in a globally naked singularity. The space-time is asymptotically
flat at ${\cal{J}}^+$ ($r\to\infty$ at fixed $U$) for fixed $z$.

\begin{figure}[!htb]
\centerline{\def\epsfsize#1#2{1 #1}\epsffile{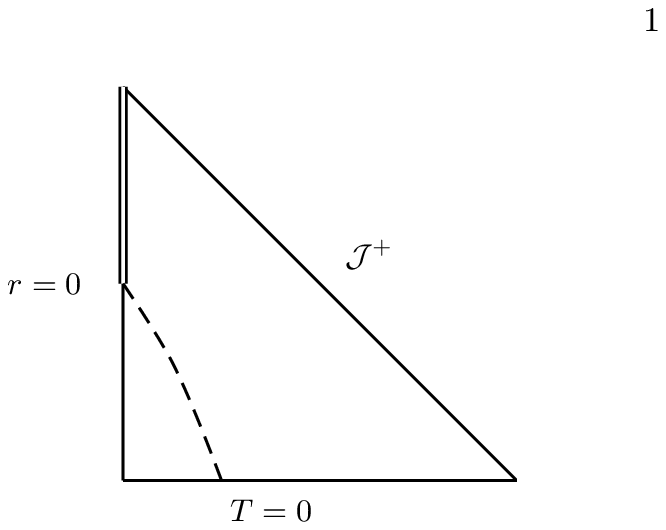}}
\caption{\label{figure7} Conformal diagram in the $U-V$ plane for
collapse of a shell of counter-rotating dust particles. The dashed
curve represent the shell; space-time is flat in the interior of
the shell. The exterior corresponds to the time-reversal of
Figures 1 or 3.}
\end{figure}

\begin{figure}[!htb]
\centerline{\def\epsfsize#1#2{1 #1}\epsffile{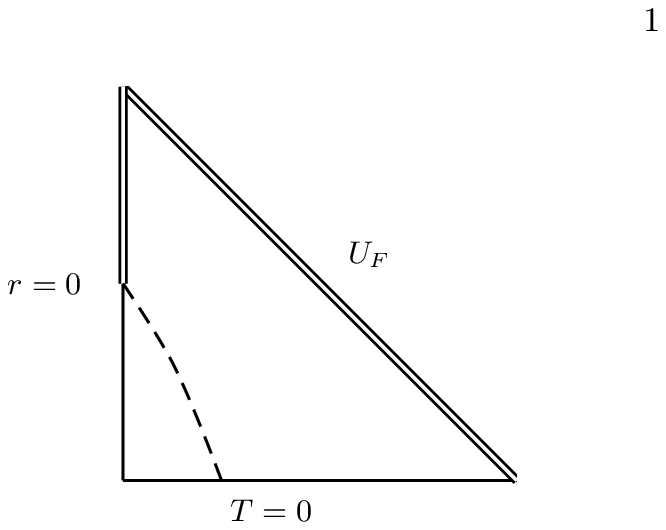}}
\caption{\label{figure8} Conformal diagram in the $U-V$ plane for
collapse of a shell of counter-rotating dust particles. The dashed
curve represent the shell; space-time is flat in the interior of
the shell. The exterior corresponds to the time-reversal of
Figures 2 or 4 with $U_F=-v_P$. The possible past singularity is
ignored.}
\end{figure}

The problem of the lack of asymptotic flatness at large $z$ can be
dealt with as follows. The space-time with line element
(\ref{linel}) can be matched smoothly across any surface
$z=$constant with the spherically symmetric space time with line
element \be
ds^2=-\exp(-\alpha(v))dudv+r^2(u,v)(d\theta^2+\sin^2\theta
d\phi^2),\label{sphere}\ee provided the matching is done across
the equator $\theta=\pi/2$ of the spherically symmetric
space-time. Doing this at two different values of $z$ and choosing
normals in the appropriate directions corresponds to replacing the
infinite cylinder with a finite cylinder bounded by hemispherical
caps \cite{BIL}. The Ricci tensor of (\ref{sphere}) has
non-vanishing components
\[ R_{vv}=-\frac{\alpha^\prime}{4r},\qquad
R_{\theta\theta}=\csc^2\theta R_{\phi\phi}=1-e^\alpha.\] The
models of Figures 5 and 6 used the boundary condition
$\alpha(0)=0$, and the energy conditions for the cylinder gave
$\alpha^\prime(v)\leq 0$. Using these, it is straightforward to
show that all energy conditions (weak, strong, dominant) are
satisfied in the hemispheres. A similar conclusion holds for the
off-shell regions of the models of Figures 7 and 8 using the
condition $\beta\leq0$ found above. The collapsing shell of matter
now has a finite cylindrical body and hemispherical caps. We take
its equation of motion to be given by the same equation
$r=r_0(T(\tau))$ in both the cylindrical and spherical regions
(this is in line with our `recycling' of the cylindrical metric
functions in the spherical regions). It is straightforward to show
that the surface energy-momentum tensor of the shell in the
cylindrical region is given by \cite{PW,GJ,Gleiser}
\[ S_{ab}=\rho\delta^\tau_a\delta^\tau_b+p_z\delta^z_a\delta^z_b
+r_0^2(p_z+\rho)\delta^\phi_a\delta^\phi_b.\] Then it transpires
that the surface energy-momentum tensor in the spherical region is
\[ S_{ab}=2\rho\delta^\tau_a\delta^\tau_b+p_zr_0^2\delta^\theta_a\delta^\theta_b
+p_zr_0^2\sin^2\theta \delta^\phi_a\delta^\phi_b.\] Imposing the
`counter-rotating dust' condition $p_z=0$ shows that the
collapsing shell is composed of dust with no tangential stresses
in the hemispheres. In particular, the energy conditions in the
hemispheres are inherited from the energy conditions in the
cylindrical region.
 Asymptotic
flatness in the hemispherical regions is guaranteed by asymptotic
flatness at fixed $z$ in the cylindrical region. The properties of
the singularities which we have examined depends only on the
geometry of the Lorentzian 2-space $z=$constant, $\phi=$constant
of (\ref{linel}). Since this 2-space is identical to the 2-space
$\theta=$constant, $\phi=$constant of (\ref{sphere}), the
singularity structure in the cylindrical region applies throughout
the entire space-time.

Finally, we note the unsuitability of these models for studying
the hoop conjecture. Since the hemispheres may be attached at any
pair of values of $z$, the resulting object may have an arbitrary
degree of prolateness, from zero (spherical) to infinity (infinite
cylinder). Thus the lack of occurrence of an apparent horizon
(i.e.\ trapped surfaces) relates only to the particular choice of
metric functions and not to the asphericity of the collapsing
object.

I am grateful to S. Gon\c{c}alves and S. Jhingan for allowing me
to read their work \cite{GJ} prior to publication.

\end{document}